\documentclass[%
 reprint,
superscriptaddress,
 amsmath,amssymb,
 aps,
prb,
floatfix,
]{revtex4-2}

\usepackage{etoolbox}
\apptocmd{\sloppy}{\hbadness 10000\relax}{}{}
\usepackage{graphicx}
\usepackage{dcolumn}
\usepackage{bm}
\usepackage{physics}
\usepackage{subfigure}
\usepackage{xcolor}
\usepackage[hidelinks]{hyperref}
\usepackage[capitalize]{cleveref}
\usepackage{changes}

\begin{document}

\preprint{APS/123-QED}

\title{Plasmons in $\mathbb{Z}_{2}$ Topological Insulators}

\author{Yuling Guan}
  \email{yulinggu@usc.edu}
 \affiliation{Department of Physics and Astronomy, University of Southern California, Los Angeles, CA 91361}
 \author{Stephan Haas}
\affiliation{Department of Physics and Astronomy, University of Southern California, Los Angeles, CA 91361}
\author{Henning Schl\"omer}
\affiliation{Department of Physics and Arnold Sommerfeld Center for Theoretical Physics (ASC), Ludwig-Maximilians-Universit\"at M\"unchen, Theresienstr. 37, M\"unchen D-80333, Germany}
\affiliation{Munich Center for Quantum Science and Technology (MCQST), Schellingstr. 4, D-80799 M\"unchen, Germany}
\author{Zhihao Jiang}
\email{zjiang22@illinois.edu}
\thanks{The research work presented in this paper started when the author was a PhD candidate in the Department of Physics and Astronomy at University of Southern California (USC), Los Angeles, CA 91361. The author thanks USC for the support during that time.}
\affiliation{Department of Materials Science and Engineering, University of Illinois at Urbana-Champaign, Urbana, IL 61801, USA}

\date{\today}

\begin{abstract}

We study plasmonic excitations in the Kane-Mele model, a two-dimensional $\mathbb{Z}_{2}$ topological insulator on the honeycomb lattice, using the random phase approximation (RPA). In the topologically non-trivial phase, the model has conducting edge states that traverse the bulk energy gap and display spin-momentum-locking. Such a state of matter is called the quantum spin hall (QSH) phase, which is robust against time-reversal (TR) invariant perturbations. We find that in the QSH phase, gapless spin-polarized plasmons can be excited on the edges of the system. The propagation of these plasmons is chiral for each individual spin component and shows spin-momentum-locking for both spin components on the same edge. Moreover, we study the effect of external magnetic fields on the gapless edge plasmons. 
Specifically, out-of-plane magnetic fields delocalize edge plasmons propagating in one direction without affecting the other one, while an in-plane magnetic field can be applied to selectively excite a specific spin-plasmon branch with proper doping or gating to the system. Our findings may have potential applications in novel plasmonic and spintronic devices. We also investigate plasmons in the Kane-Mele model on a finite-sized diamond-shaped nanoflake and observe low-energy plasmons circulating the boundary of the material.       
\end{abstract}

\maketitle

\section{Introduction}\label{sec:Intro}

The experimental observation of the integer quantum Hall effect (IQHE) \cite{IQHE} has led to a fundamentally novel way of classifying electronic states via topological invariants. In the pioneering work of Thouless, Kohmoto, Nightingale, and den Nijs \cite{tknn82} (TKNN), the topological invariant for the IQHE was identified as an integer called the Chern number. Here, non-trivial topology is attributed to a broken time reversal (TR) symmetry \cite{h88}. In the context of TR-invariant systems, where the Chern number vanishes, Kane and Mele \cite{km05, km05_2} introduced another $\mathbb{Z}_{2}$ topological classification which distinguishes the quantum spin Hall (QSH) phase from the trivial insulating phase. The QSH phase is characterized by spin-momentum-locked topological surface states traversing the bulk energy band gap. Kramers theorem~\cite{kramers1930} guarantees the degeneracy of energy bands at TR invariant momenta (TRIM) and forbids gap opening on the surface from any TR-invariant perturbation~\cite{vanderbilt2018}.

While the topological electronic properties have been extensively studied in the past few decades \cite{Zhang2005, Bernevig2006, Bernevig2006HgTe, Konig2007, Novoselov2007, Fu2007, Fu2007TI_IS, Hsieh2008, Chen2009_3DTI, Kuroda2010, Gong2019, Noguchi2019}, the behavior of  collective excitations, such as plasmons \cite{Karch2011, Okada2013, Schutky2013, Efimkin2012, Yuan2017EXP, Dubrovkin2017EXP, Jin2017, DiPietro2013EXP, Venuthurumilli2019EXP, Lai2014, Ginley2018EXP} and magnons~\cite{Ruckriegel2018, Malki2019, Mook2014} in topologically non-trivial systems is drawing increasing attention as well. Our previous studies have found that plasmonic excitations in the Su-Schrieffer-Heeger (SSH) model~\cite{jrgh20,gjh21,sjh21} and the Haldane model~\cite{sjh21} are clearly affected by the underlying electron topology, and consequently display characteristic behavior such as gapless spectra, localized modes on the sample surface, and robustness against appreciable disorder. While these models consider spinless electrons, in $\mathbb{Z}_{2}$ topological insulators (TIs) it is necessary to treat electrons as spinors due to possible spin-mixing couplings. In fact, plasmons in such systems have been studied on surfaces of two-dimensional (2D)~\cite{Lai2014} and three-dimensional (3D)~\cite{Lai2014, Karch2011, Efimkin2012, Efimkin2021} TIs. Due to the spin-momentum-locking of the conducting edge states in $\mathbb{Z}_{2}$ TIs, these surface plasmons are spin polarized along certain directions. For this reason, they are called "spin-plasmons", with  novel applications in plasmonics and spintronics~\cite{Appelbaum2011}. Moreover, spin-polarized plasmons were also reported in the two-dimensional electron gas that are observed in Raman-scattering experiments~\cite{apvf14}, and in spin-polarized graphene that shows long lifetime dispersing in its undamped regime~\cite{av15}. However, we are not aware of any studies of how such surface spin-plasmons can be engineered and tuned by e.g. the application of electromagnetic fields. Considering the potentially significant applications of spin-plasmons, we believe it is useful to report here our studies on controlling surface spin-plasmons. 

In this paper, we consider the Kane-Mele model~\cite{km05}, a paradigmatic model of 2D $\mathbb{Z}_{2}$ TIs, on a ribbon structure. In its TR-symmetric QSH phase, we observe spin-polarized gapless edge plasmons displaying chirality, spin-momentum-locking and spin-edge-locking that are inherited from the topology and symmetry of the underlying single-electron states. We find that these plasmons are sensitive to external magnetic fields applied in different directions, which can therefore be used as a practical technique to control them, although the field breaks the TR symmetry such that some topological features of these plasmons are lost. Specifically, we show that an out-of-plane magnetic field can be applied to tune surface plasmon localization, and an in-plane magnetic field can be used to selectively excite plasmons with certain spin polarization. Moreover, we calculate the plasmonic excitation spectrum in real-space in a diamond-shaped Kane-Mele model and observe edge plasmons circulating the sample. The single-electron topological properties of this model have recently been investigated in \cite{rqq20}.

The remainder of this paper is organized in the following way. In Sec.~\ref{sec:Method}, we introduce the random phase approximation (RPA) method applied to the ribbon structure for calculating the dielectric response and plasmonic excitations. In Sec.~\ref{sec:Result}, we present  detailed results of plasmonic excitations in the Kane-Mele model. Plasmons in the original Kane-Mele ribbon structure are firstly introduced, followed by two variant scenarios when the original system is exposed to an external magnetic field applied in the out-of-plane direction and the in-plane direction. This is followed by a discussion of plasmons in real space on a diamond-shaped Kane-Mele model with open boundaries. Finally, we conclude our study, proposing potential applications of these results and give future research directions in Sec.~\ref{sec:conclusion}.

\section{METHOD}\label{sec:Method}

In order to study plasmonic excitations, we calculate the dielectric function within the random phase approximation (RPA). The RPA dielectric function has traditionally been introduced in momentum space for the homogeneous electron gas~\cite{pb52,bp53}, as well as recently been formulated fully in real space to study non-translational-invariant systems~\cite{wcjtwmn15,wvky18,jrgh20,gjh21}. In this paper, we mainly focus on ribbon structures, which are periodic in the longitudinal direction while finite-sized in the transverse direction (Fig.\ref{fig:Structure}(b)). Therefore, we adopt a combined treatment of both the real- and momentum space as detailed below. 

Generically, as indicated in Fig.~\ref{fig:Structure} (b), the unit cell spans the entire width of the ribbon (along $X$-direction) and is repeated periodically along the $Y$-direction. We denote the position and the spin of the $i$-th atom in the unit cell as $\vec{\tau}_i$ and $\vec{\sigma}_i$. By Fourier transformation along the $Y$-direction, we can construct a tight-binding basis $\{\ket{\mu} \equiv \ket{\vec{\tau}}\otimes\ket{\sigma}, \vec{\tau}=\vec{\tau}_1,\vec{\tau}_2,\dots,\vec{\tau}_N \text{ and } \sigma=+,-\}$ for each $k_y$, supposing there are $N$ atoms in the unit cell. Using this, we can express the non-interacting polarization function as a matrix, whose element indexed by $\mu$ and $\mu'$ is given by~\cite{wvky18}
\begin{equation}
\begin{aligned}
[\chi_0(\omega, q_y)]_{\mu\mu'}=&\frac{1}{V}\,\sum_{k_y,n,n'}\frac{f(E^{k_y}_n)-f(E^{k_y+q_y}_{n'})}{\omega+i\gamma+E^{k_y}_n-E^{k_y+q_y}_{n'}}\\&\cross\psi_{n\mu}^{k_y} (\psi_{n\mu'}^{k_y})^*(\psi_{n'\mu}^{k_y+q_y})^*\psi_{n'\mu'}^{k_y+q_y},
\end{aligned}
\label{eq:susceptibility}
\end{equation}
where $\omega$ is the frequency and $q_y$ is the momentum transfer in $Y$-direction. $V$ is the volume of the unit cell in the periodic direction. $E^{k_y}_n$ and $\psi_{n\mu}^{k_y}$ are the eigenenergy and the $\mu$-th component of the eigenvector of the state with band index $n$ and momentum $k_y$. They can be computed from diagonalization of the tight-binding Hamiltonian $H(k_y)$ written in the same basis introduced above. $f(\cdot)$ is the Fermi-Dirac distribution function, whose zero temperature limit is applied in all calculations in this paper. Moreover, we introduce a finite broadening parameter $\gamma$ which is set to be $0.01\ \text{eV}$.

On the other hand, we consider electron-electron Coulomb interaction which is also Fourier transformed along the $Y$-direction, namely, $V(q_y)$. Using the same basis and assuming that the Coulomb interaction only depends on the real-space coordinates of two sites \footnote{We want to clarify that this is a simplified assumption. The density-density Coulomb interaction can depend on electron spin especially when the real-space distance is small, however, this is out of the scope of the current paper and will be considered in future works.}, we have 
\begin{equation}
V_{\mu\mu'}(q_y)=\left\{
\begin{aligned}
& 2e^2*K_0(|q_y||x - x'|)/\kappa , & \text{if}\; \vec{\tau} \neq \vec{\tau'}, \\
& U_0/\kappa, & \text{if}\; \vec{\tau} = \vec{\tau'}.
\end{aligned}
\right.
\label{eq:Coulomb}
\end{equation}
Here $K_0(\cdot)$ is the zeroth modified Bessel function of the second kind. This form of the Coulomb interaction is referred from \cite{s93}. For the onsite Coulomb interaction, we use a parameter of $U_0=17.38\,eV$ \cite{jrgh20} to avoid divergence. Additionally, $\kappa$ is the dielectric constant of the background medium which is set to be $\kappa=1$ (the vacuum) unless otherwise specified.  

We can then derive the dielectric matrix within RPA as
\begin{equation}
\epsilon_\text{RPA}(\omega, q_y) = \text{I} - V(q_y)\,\chi_0(\omega, q_y),
\end{equation}
form which we can further extract the electron energy loss spectrum (EELS) by 
\begin{align}
    \text{EELS}(\omega,q_y)=\max_i \bigg \{ -\text{Im}\ \left[\frac{1}{\epsilon_i(\omega,q_y)}\right] \bigg \}, \label{eq:EELS}
\end{align}
where $\epsilon_i(\omega, q_y)$ is the i-th eigenvalue of $\epsilon_\text{RPA}(\omega, q_y)$. Plasmonic excitations are identified as peaks in EELS$(\omega, q_y)$. Meanwhile, the real-space charge distribution patterns of plasmon modes can also be obtained utilizing the right eigenvector of the dielectric matrix $\epsilon_\text{RPA}(\omega, q_y)$. The method is similar to the one reported in~\cite{jrgh20} for the full real-space RPA calculation. Suppose that $\epsilon_m(\omega_p, q_y)$ is the selected eigenvalue indicating a plasmon mode at $\omega_p$, then the real-space charge distribution pattern of this mode can be obtained by
\begin{equation}
\label{eq:ChargeDensity}
\rho_0(\omega_p,q_y)=\chi_0(\omega_p,q_y)\xi_{m}(\omega_p,q_y),
\end{equation}
with $\xi_{m}(\omega_p,q_y)$ is the right eigenvector corresponding to $\epsilon_m(\omega_p, q_y)$. In fact, we can also obtain the second EELS (2nd-EELS) by selecting the second maximum in Eq.~\eqref{eq:EELS}, which will provide information about degeneracy of plasmon modes. In this paper, we will always consider the maximum EELS unless otherwise specified. 

\section{Models and Results}\label{sec:Result}

\subsection{Chiral edge spin-plasmons in the Kane-Mele ribbon structure}
\begin{figure}[htb]
\centering
\begin{minipage}{\linewidth}
\centering
\includegraphics[width=\linewidth]{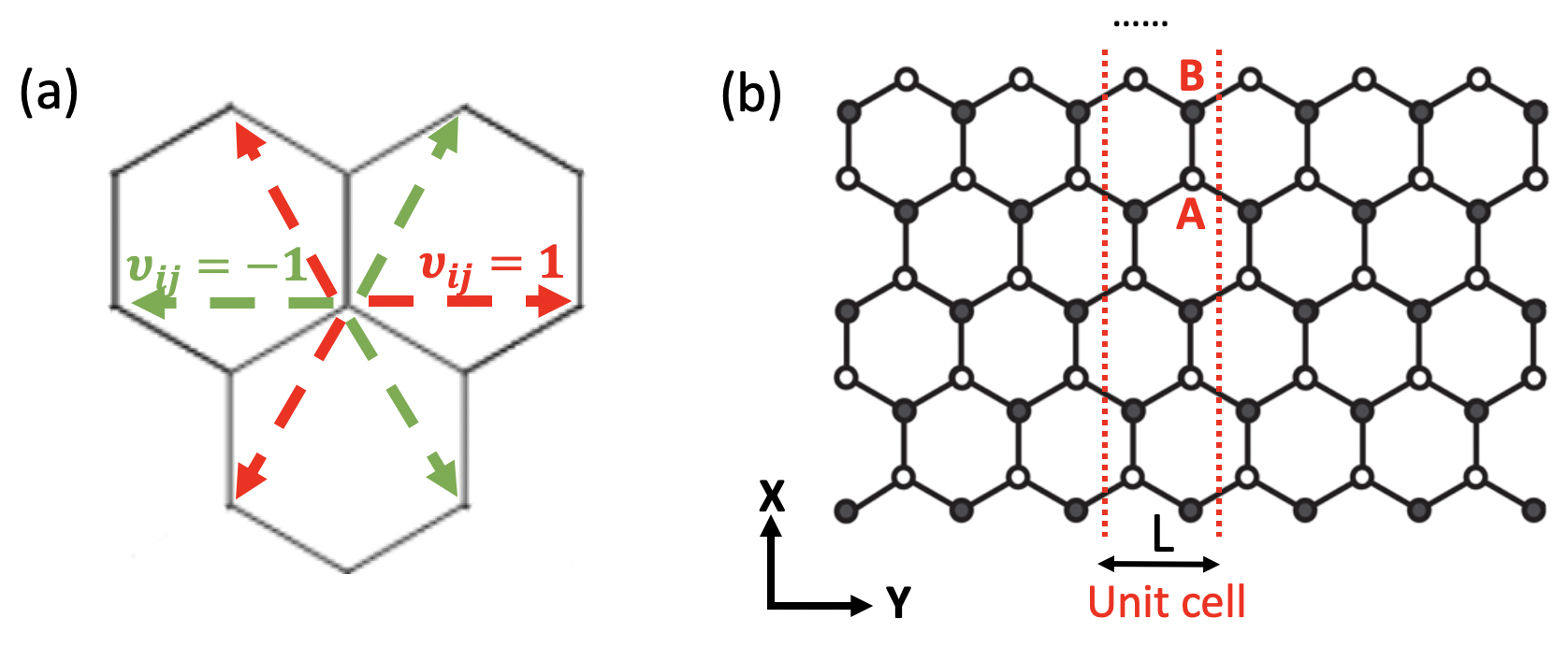}
\end{minipage}
\caption{(a) Determination of the sign $\nu_{ij}$ for the spin-orbit coupling term in the Kane-Mele model on a  Honeycomb lattice. (b) Illustration of Kane-Mele model on a ribbon structure with zigzag terminations. Two atomic species A and B are non-equivalent with a non-zero onsite energy $t_v$. Two dashed red lines define a unit cell that is repeated along the $Y$-direction with period $L$.}
\label{fig:Structure}
\end{figure}

\begin{figure}[htb]
\centering
\begin{minipage}{\linewidth}
\centering
\includegraphics[width=\linewidth]{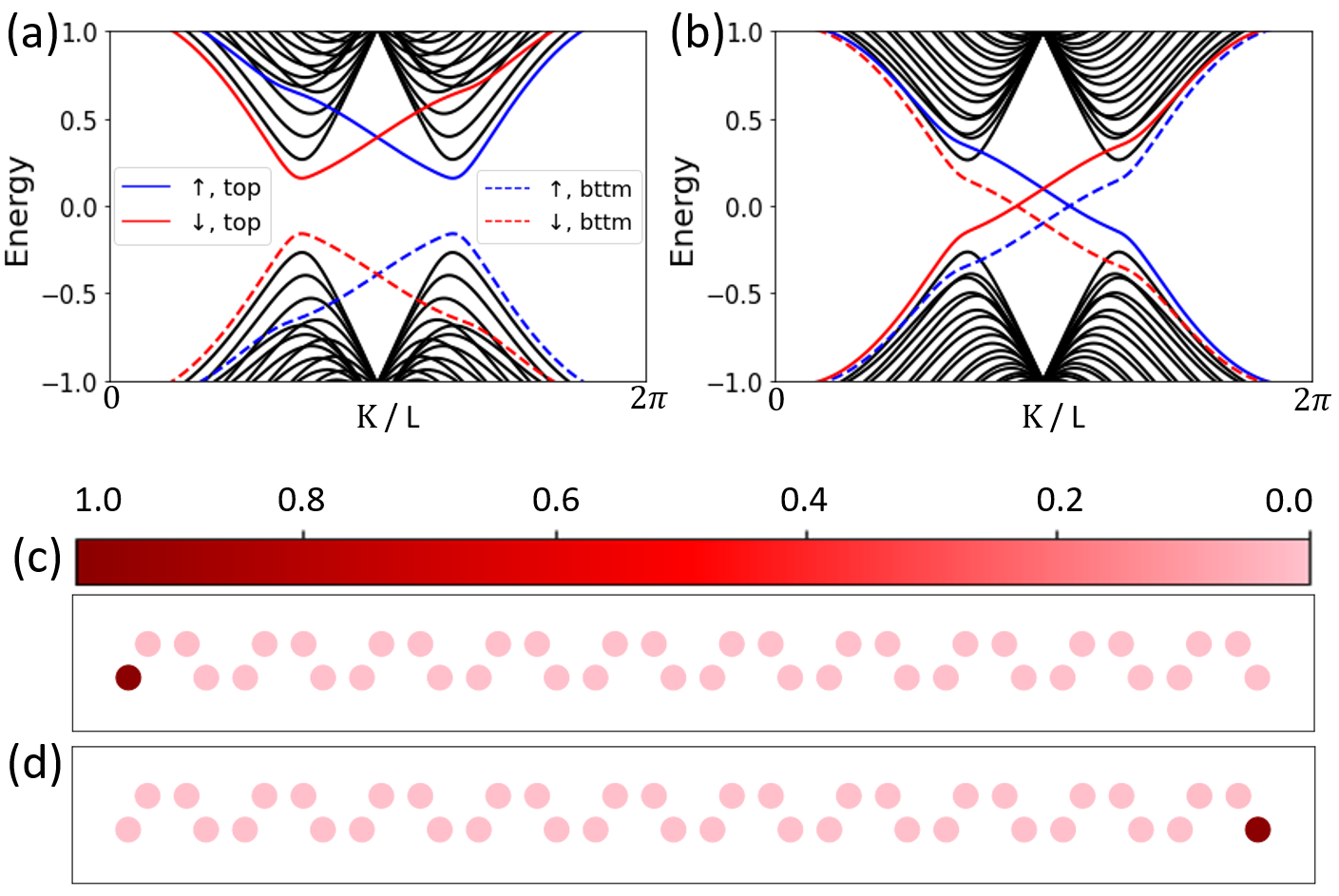}
\end{minipage}
\caption{Electronic band structure of the Kane-Mele model on a ribbon with zigzag terminations. (a) Trivial insulating phase with $t_{\mathrm{SO}} = 0.05t$ and $t_v = 0.4t$. (b) Quantum Spin Hall (QSH) phase with $t_{\mathrm{SO}} = 0.05t$ and $t_v = 0.1t$. Conducting edge states traversing the bulk energy band gap are observed. Two spin components are represented by blue and red bands; two edges are represented by solid and dashed bands. (c) and (d) are modular square of electronic wavefunctions of the two spin-up edge states at the momentum $K = 1.01\pi/L$. They are strongly localized on two opposite edges.}
\label{fig:KM_band}
\end{figure}

The Kane-Mele model on a honeycomb lattice is given by the Hamiltonian~\cite{km05}
\begin{equation}
\begin{aligned}
H = &t \sum_{\langle i, j\rangle} c_{i}^{\dagger} c_{j}+i t_{\mathrm{SO}} \sum_{\langle\langle i, j\rangle\rangle} \nu_{i j} c_{i}^{\dagger} s^{z} c_{j} \\
    &+i t_{R} \sum_{\langle i, j\rangle} c_{i}^{\dagger}\left(\mathbf{s} \times \hat{\mathbf{d}}_{i j}\right)_{z} c_{j} +t_{v} \sum_{i} \xi_{i} c_{i}^{\dagger} c_{i}.
\end{aligned}
\label{Eq:Hamiltonian}
\end{equation}
Here $c_{i}^{\dagger}=\left(c_{i \uparrow}^{\dagger}, c_{i \downarrow}^{\dagger}\right)$ and $c_{i}=\left(c_{i \uparrow}, c_{i \downarrow}\right)$ are electron creation and annihilation operators. The first term is the hopping between nearest neighbor sites denoted as $\langle i, j\rangle$. The second term describes the mirror symmetric intrinsic spin-orbit coupling involving next nearest neighbor sites $\langle\langle i, j\rangle\rangle$. The sign $\nu_{ij}$ depends on the orientation of the hopping as illustrated in Fig.~\ref{fig:Structure}(a). Mathematically, it can be determined by $\nu_{i j}=(2 / \sqrt{3})\left(\hat{\mathbf{d}}_{1} \times \hat{\mathbf{d}}_{2}\right)_{z}$,
where $\hat{\boldsymbol{d_{1}}}$ and $\hat{\boldsymbol{d_{2}}}$ are unit vectors along the two bonds that traverse from the $j$th site to the $i$th site. $s_z$ is the diagonal Pauli matrix. $t_{\mathrm{SO}}$ is the amplitude for spin-orbit interaction and can be determined by second-order perturbation theory calculations~\cite{hgb06, mhsskm06, yyqzsf07, bt07}. The third term is the Rashba coupling between nearest neighbor sites. It introduces an off-diagonal spin mixing term that can arise from applying a perpendicular electric field or placing the sample on a substrate. Its amplitude $t_R$ can be experimentally measured~\cite{vssbvrmr08, ttvjpjv08}. The last term is a staggered sublattice potential ($\xi_i=\pm1$) on site A and B [Fig.~\ref{fig:Structure}(b)], which breaks the in-plane twofold rotation symmetry for non-zero $t_v$. In a ribbon structure of Fig.~\ref{fig:Structure}(b), it breaks the symmetry between two edges as the top and bottom edge are terminated at different atomic species. In our numerical calculations, we set the nearest neighbor hopping parameter $t=1\ \text{eV}$ and all other energies are scaled in the unit of $t$. We also set $t_R=0$, such that spin-up and spin-down electrons are originally decoupled.

We focus on the Kane-Mele model on the ribbon structure with zigzag terminations as illustrated in Fig.~\ref{fig:Structure}(b). For $t_R=0$, the model has a bulk energy band gap $E_g = \left|6 \sqrt{3} \lambda_{\text {SO }}-2 \lambda_{v}\right|$~\cite{km05} that vanishes when $\lambda_{v}=3 \sqrt{3} \lambda_{\mathrm{SO}}$. For $\lambda_{v}>3 \sqrt{3} \lambda_{\mathrm{SO}}$, the system is in the trivial insulating phase, while for $\lambda_{v}<3 \sqrt{3} \lambda_{\mathrm{SO}}$ it is in the topologically non-trivial quantum spin Hall (QSH) phase~\cite{km05}. This distinction is illustrated via the electronic single particle band structure in Fig.~\ref{fig:KM_band}. In Fig.~\ref{fig:KM_band}(a), we choose $t_{\mathrm{SO}} = 0.05t$ and $t_v = 0.4t$ (the trivial insulating phase) and observe no edge states traversing the bulk energy gap. In Fig.~\ref{fig:KM_band}(b) (with $t_{\mathrm{SO}} = 0.05t$ and $t_v = 0.1t$, i.e. the system is in the QSH phase), we observe conducting edge states crossing the energy gap. For any energy within the bulk band gap, there is one TR edge-states pair on each edge, indicating a non-trivial (odd) $\mathbb{Z}_{2}$ topological order. The conducting edge states for spin-up and spin-down electrons on each edge propagate in opposite directions, which is a characteristic property called ``spin-momentum-locking" in the QSH phase. We plot the modular square of the wave functions $|\psi|^2$ for two spin-up electronic edge states at zero energy (Figs.~\ref{fig:KM_band}(c) and (d)), confirming that they are localized on two opposite edges. The same properties apply for spin-down electronic edge states. Due to the non-zero $t_v$ in the calculation, energy spectra of conducting edges states on the top edge (solid lines) and the bottom edge (dashed lines) are shifted.        

\begin{figure}[htb]
\centering
\begin{minipage}{\linewidth}
\centering
\includegraphics[width=\linewidth]{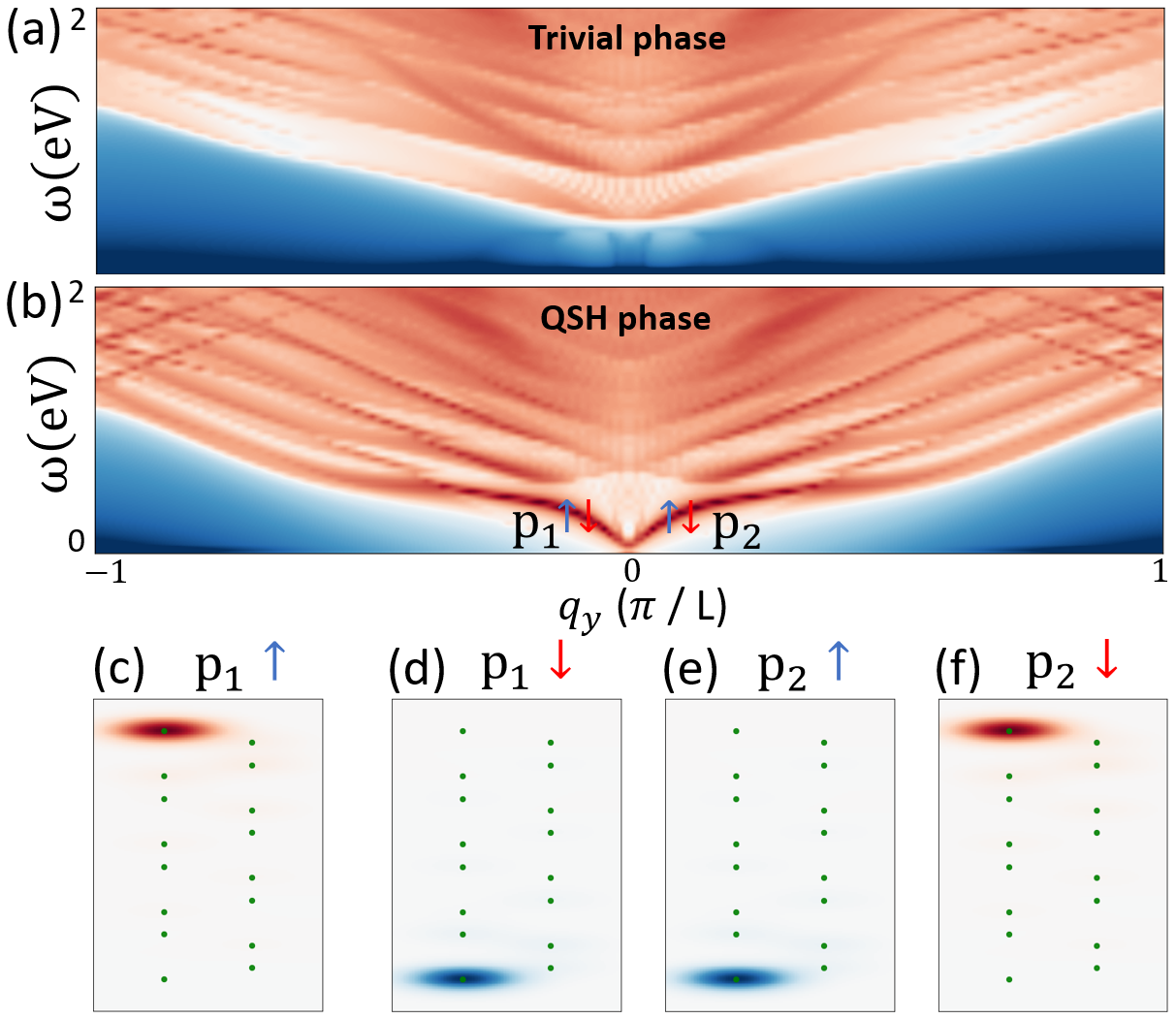}
\end{minipage}
\caption{Electron energy loss spectra (EELS) of the Kane-Mele model on a ribbon structure in (a) the trivial insulating phase and (b) the in QSH phase. Gapless spin-plasmons are observed on the edges of the QSH system. (c-f) show real-space charge modulation patterns of the spin-up plasmon at $\mathrm{P}_1$, the spin-down plasmon $\mathrm{P}_1$, the spin-up plasmon at $\mathrm{P}_2$ and the spin-down plasmon at $\mathrm{P}_2$. Red and blue in the modulations represent the positive and negative charges.}
\label{fig:KM_EELS}
\end{figure}

We now analyze the plasmonic excitations in the Kane-Mele ribbon structure. Firstly, we show that \textit{gapless edge plasmons} can be excited in the QSH phase, but not in the trivial insulating phase. In Figs.~\ref{fig:KM_EELS}(a) and (b), we plot the plasmon dispersions for both phases. We observe that in the QSH phase, gapless low-energy plasmons emerge due to the existence of the conducting edge states observed in the single-electron band structure (see Appendix.~\ref{app:Topo_origin} for a detailed analysis of the relation between both). These plasmons are localized on the edges of the ribbon, as confirmed by the real-space charge distribution patterns of some typical modes shown in Figs.~\ref{fig:KM_EELS}(c-f). In contrast, gapless edge plasmons are absent in the trivial insulating phase. Therefore, they can be regarded as a  topological signature of collective excitations in QSH insulators. 

Secondly, we find that gapless edge plasmons are spin-resolved and inherit \textit{chirality} as well as \textit{spin-momentum-locking} from the QSH electronic structure. \textit{Chirality} of the gapless edge plasmons can be detected by focusing on each individual spin component. In the spin-decoupled QSH regime ($t_R=0$), each spin component behaves effectively as a non-trivial Chern insulator displaying chiral edge current flow. We analyze in detail the inheritance of chirality from the electronic structure to the plasmonic excitations for the spin-up component, while the spin-down component can be understood in the exactly same way. The spin-up edge current propagates uni-directionally with positive (negative) momentum on the bottom (top) edge, as indicated by the positive (negative) slope of the dashed (solid) blue edge-states energy dispersion in Fig.~\ref{fig:KM_band}(b). This property is inherited by the plasmons in the sense that the spin-up edge plasmons with positive (negative) momenta $q_y$ can only stay on the bottom (top) edge, as indicated by the mode profile of ``$P_2\uparrow$" in Fig.~\ref{fig:KM_EELS}(e) (``$P_1\uparrow$" in Fig.~\ref{fig:KM_EELS}(c)). This can be understood by noticing that for the positive (negative) momentum transfer $q_y$, spin-up electronic transitions can only occur within the dashed (solid) blue band in Fig.~\ref{fig:KM_band}(b), along which all electronic states are localized on the bottom (top) edge of the ribbon. It is true that electronic transitions with positive (negative) $q_y$ can also occur within the solid (dashed) red band. However, this band includes spin-down electronic states localized on the top (bottom) edge of the ribbon, which have negligible interfere with the dashed (solid) blue band of spin-up electrons on the bottom (top) edge due to their large real-space separation. On the other hand, \textit{spin-momentum-locking} of gapless edge plasmons can be inferred by focusing on each individual edge. On the top edge, spin-up plasmons can only be excited with negative momentum $q_y$ (Fig.~\ref{fig:KM_EELS}(c)), whereas spin-down plasmons are restricted to positive momenta. On the bottom edge, we find the opposite behavior for the two spin components. The reason for the spin-momentum-locking of gapless edge plasmons can also be understood by analyzing the restricted transitions within each conducting electronic bands in Fig.~\ref{fig:KM_band}(b).

Lastly, if we excite gapless edge plasmons at a specific momentum $q_y$, TR symmetry and spin-momentum-locking in the QSH phase transform into a new special property of these plasmons, which we will call  \textit{spin-edge-locking}. For instance, using an external excitation field with a specific positive $q_y$, spin-up plasmons can only be excited on the bottom edge (Fig.~\ref{fig:KM_EELS}(e)), while spin-down plasmons can only be excited on the top edge (Fig.~\ref{fig:KM_EELS}(f)). This  property opens up possibilities for plasmon engineering using these gapless edge spin-plasmons by gating, doping or substrate manipulation on the edges. For instance, the local environments on the two edges can be slightly biased such that spin-up edge plasmons and spin-down edge plasmons can be energetically separated. We can then selectively excite the spin-up or spin-down component individually by choosing proper energy-momentum combination $(\omega, q_y)$.

\subsection{Tuning edge plasmons via external magnetic fields}


In this section, we study the Kane-Mele model in the presence of external magnetic field. The original model Hamiltonian is modified by adding a Zeeman energy term induced by the external magnetic field $\mathbf{B}$, which is given here as
\begin{equation}
H^\text{Z}=H_\text{KM} + \sum_{i} c_{i}^{\dagger} \mathbf{B} \cdot \boldsymbol{s} c_{i}.
\label{Eq:Hamiltonian-ZM}
\end{equation}
$H_\text{KM}$ refers to the Kane-Mele model Hamiltonian from Eq.~\ref{Eq:Hamiltonian}. As we shall see in the following, although applying magnetic field breaks the TR symmetry, it can be useful to tune the edge plasmons excited in the Kane-Mele model. The tuning effect depends on the direction of applied Zeeman field. 

\subsubsection{Control of edge plasmons localization via $\mathbf{B_z}$ field}

\begin{figure}[htb]
\centering
\begin{minipage}{\linewidth}
\centering
\includegraphics[width=\linewidth]{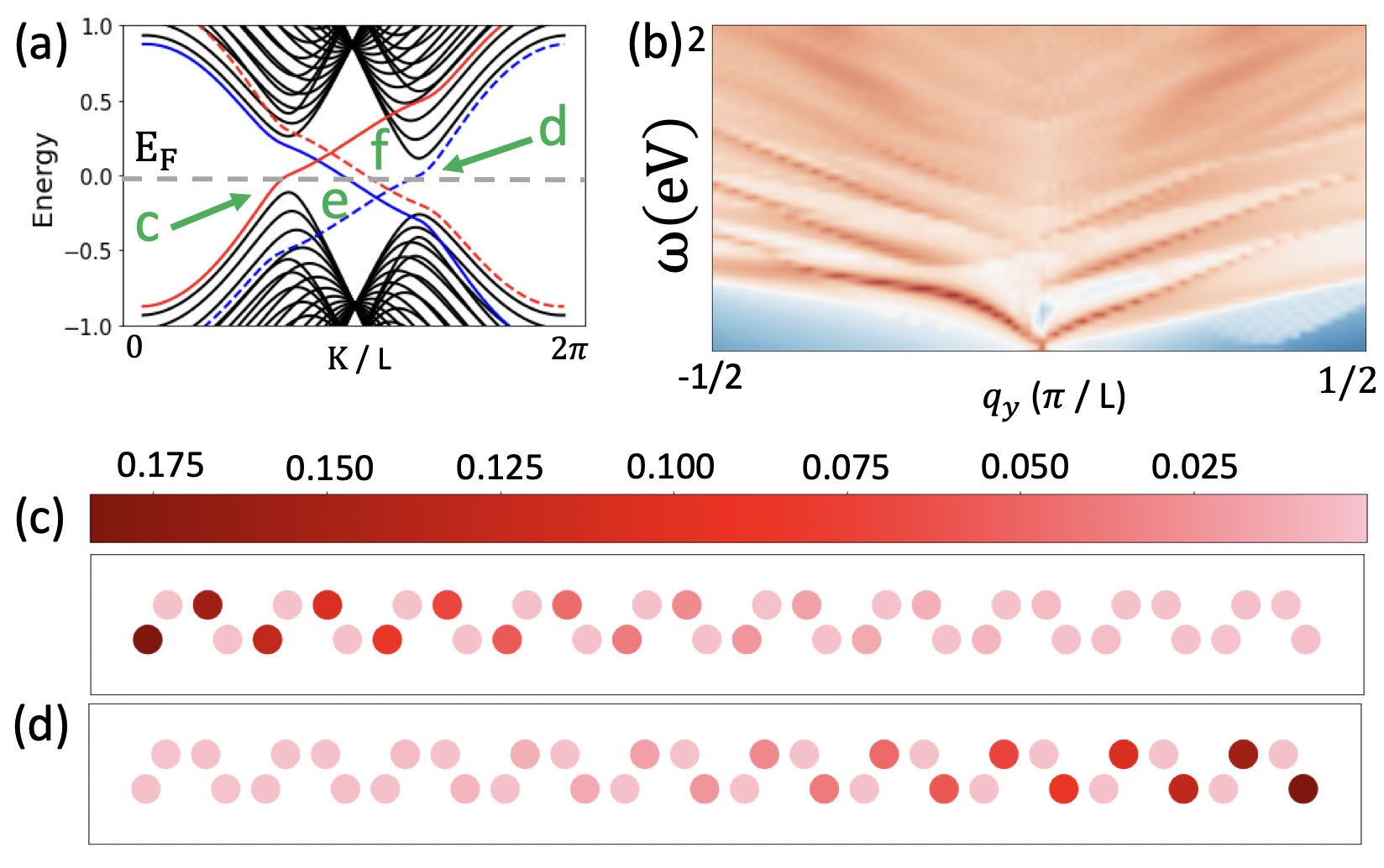}
\end{minipage}
\caption{(a) Electron band structure, and (b) plasmon dispersions, of the QSH Kane-Mele ribbon structure in presence of an out-of-plane magnetic field $\mathbf{B_z}.$ (c) and (d) are modular square of electronic eigenstates at ``c" and ``d" points marked in the electron band structure (a).}
\label{fig:Zeeman_band}
\end{figure}


We first consider the external magnetic field applied in the out-of-plane direction, namely $\mathbf{B_z} = B_z\hat{\mathbf{z}}$. In Fig.~\ref{fig:Zeeman_band}(a), we show the electronic band structure of the QSH Kane-Mele ribbon when $B_z=0.15t$. There are three main effects to be addressed due to the presence of the magnetic field. Firstly, introducing $\mathbf{B_z}$ does not open a gap on the edge of the system. For each spin component, conducting edge states crossing over the bulk band gap remain. Therefore, gapless edge plasmons are preserved, which is shown by the EELS in Fig.~\ref{fig:Zeeman_band}(b). Secondly, the field breaks the TR symmetry on each edge because the system is magnetized. This will correspondingly break TR symmetry in the corresponding plasmon branches, as can be seen by the asymmetric dispersion for $q_y \lessgtr 0$ in Fig.~\ref{fig:Zeeman_band}(b). In the previous section in contrast, where the model is non-magnetic, the dispersion is symmetric around $q_y = 0$, indicating presence of TR symmetry (Fig.~\ref{fig:KM_EELS}(a)). 
Such an asymmetric property of plasmons may find potential applications in non-reciprocal plasmonic and spintronic devices. Thirdly, although the $\mathbf{B_z}$ field does not destroy gapless edge plasmons, the real-space localization of these edge plasmons can be affected. Particularly, significant delocalization for plasmons with small positive $q_y$ is expected, because they are dominated by single electron states near ``c" and ``d" points (marked in Fig~\ref{fig:Zeeman_band}(a)) crossed by the Fermi level. As we can see, these two points are very close to bulk bands, such that their wave functions are delocalized (cf. Figs.~\ref{fig:Zeeman_band}(c) and (d), Figs.~\ref{fig:KM_band}(c) and (d)). 
In contrast, we do not expect significant delocalization effects to appear for plasmons with small negative $q_y$, because states ``e'' and ``f'' are still deep in the gap.

\begin{figure}[htb]
\centering
\begin{minipage}{\linewidth}
\centering
\includegraphics[width=\linewidth]{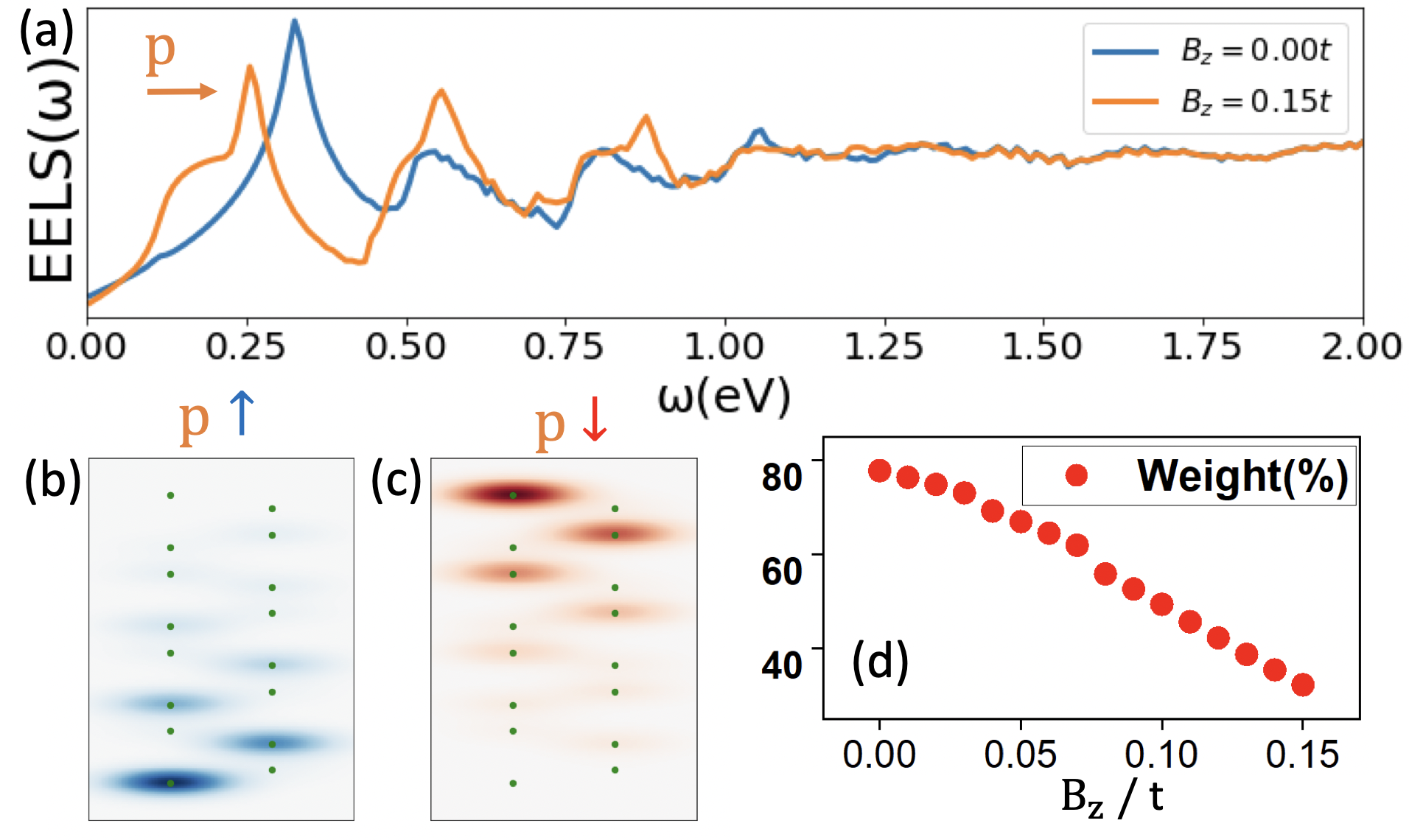}
\end{minipage}
\caption{(a) EELS of the QSH Kane-Mele ribbon structure evaluated along the momentum transfer $\text{q} = \pi/8\text{L}$ with and without an external magnetic field $\mathbf{B_z}=0.15t\hat{\mathbf{z}}$. Spin-orbit interaction for the numerical calculation here is set to be $t_{\mathrm{SO}}=0.05\,t$. (b) and (c) are real-space charge modulation patterns of two spin-plasmons excited at ``P" in the presence of the external field $\mathbf{B_z}$. They are delocalized into the bulk. (d) shows a decreasing weight of charges on the edge for the plasmon mode ``P" with increased strength of the field $\mathbf{B_z}$.}
\label{fig:Bz_EELS_singkleK}
\end{figure}

We investigate the delocalization effect on positive $q_y$ edge plasmons in more detail. Here we select modes along $q_y=\pi/8L$ and plot the EELS as a function of $\omega$, namely, $\text{EELS}(q_y=\pi/8L, \omega)$. This is given in Fig.~\ref{fig:Bz_EELS_singkleK}(a) for both $B_z=0$ (no magnetic field) and $B_z=0.15t$. The first mode in each spectrum is the edge plasmon. As we can see, introducing $B_z$ changes the excitation energy of the edge plasmon, an effect induced by the changed electronic band structure. More interestingly, the mode localization in the real space is strongly affected by $B_z$. In Figs.~\ref{fig:Bz_EELS_singkleK}(b) and (c), we show the real-space charge modulation patterns of spin-up and spin-down plasmons excited at ``p". While the ``spin-edge-locking" character remains, both modes strongly delocalize compared to their counterparts in Figs.~\ref{fig:KM_EELS}(e) and (f) without external magnetic fields. In fact, localization of edge plasmons can be gradually changed by tuning the field strength $B_z$. In Fig.~\ref{fig:Bz_EELS_singkleK}(d), we show how the weight of charges localized on the edge decreases with an increased value of $B_z$ in the range of $0$ to $0.15t$. This observation indicates a way to control edge plasmons localization via an external $\mathbf{B_z}$ field. We note that the tuning effect depends on the original bulk band gap of the system. If the original bulk band gap is large, a small external field $B_z$ will not have a big influence on the plasmon spectrum and mode localization. Detailed analysis for this case is given in Appendix~\ref{app:large_gap_Bz}.        

\subsubsection{Selective excitation of spin-polarized plasmons with applied $\mathbf{B_x}$ field}

\begin{figure}[htb]
\centering
\begin{minipage}{\linewidth}
\centering
\includegraphics[width=\linewidth]{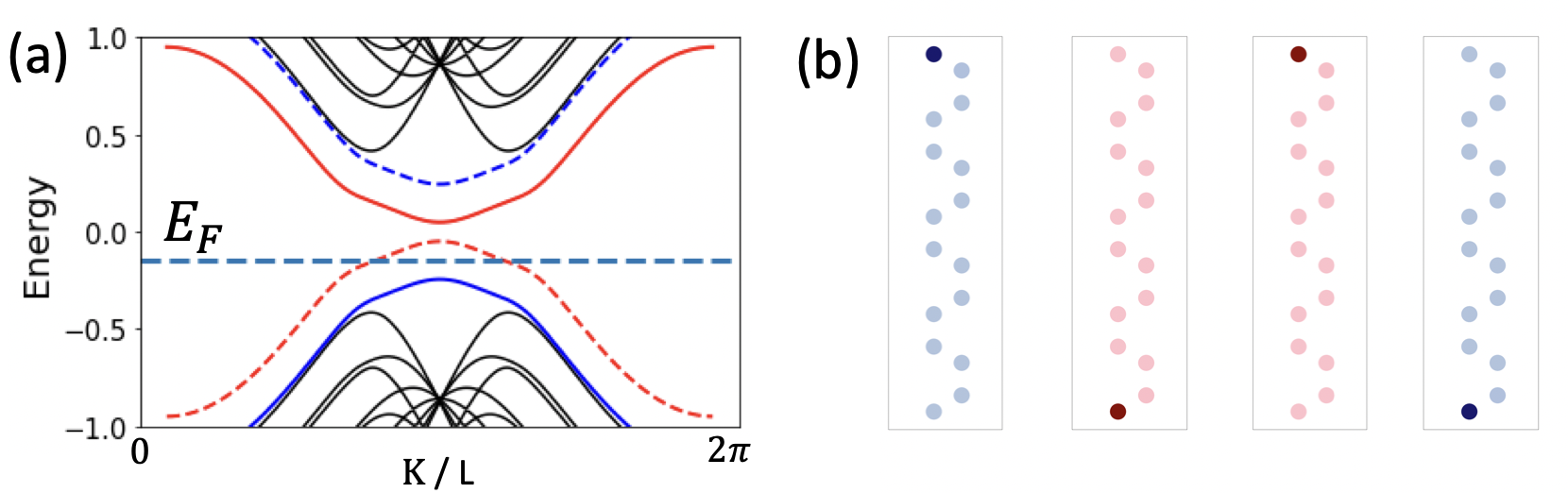}
\end{minipage}
\caption{Electronic band structure of the Kane-Mele ribbon structure in the presence of an in-plane magnetic field $\mathbf{B_x}=0.15\,t\,\mathbf{\hat{x}}$. An energy gap is opened on each edge due to the spin-flip TR-broken term. Four bands in the middle correspond to spin-up and spin-down electronic states localized on the top and bottom edges, represented in the same way as in Fig.~\ref{fig:KM_band}. (b) Modular square of electronic eigenstates of the middle four bands at $K = 1.01\pi/L$. The sub-figures are ordered from low energy to high energy. Strong localization character is still preserved in the presence of $\mathbf{B_x}$ field.}
\label{fig:Bx_band}
\end{figure}

We now turn to the QSH Kane-Mele model in the presence of an in-plane magnetic field $\mathbf{B_x} = B_x\hat{\mathbf{x}}$ applied in $x$-direction. This induces an off-diagonal on-site spin-flip term that violates the time-reversal symmetry, which opens a gap on each edge of $E_g=2B_x$ (Fig.~\ref{fig:Bx_band}(a)). For strong enough $B_x$, the model is completely gapped at zero energy. Gapless edge plasmons will therefore vanish if we keep the Fermi level at $E_F=0\text{ eV}$. However, the localization character of electron eigenstates on these bands is still preserved, as long as these bands still extend deeply in the bulk energy gap. This can be confirmed the modular square of the four electronic eigenstates at $k=1.01\pi/L$ plotted in Fig.~\ref{fig:Bx_band}(b). 

The above property of the electronic structure in the presence of $\mathbf{B_x}$ implies a way to selectively excite gapless edge plasmons with specific spin polarization. As we can see, due to the gap opening on each edge, the spin-up electron band and the spin-down electron band are energetically separated. We can tune the Fermi level by doping \cite{mfdgl09,wlwcwd06} or gating \cite{cmwg94} the system such that $E_F$ crosses one band of a specific spin polarization, as illustrated in Fig.~\ref{fig:Bx_band}(a). In this case, the spin-down electron band on the bottom edge (the dashed red line) becomes conducting. Therefore, gapless edge plasmons can just be excited for spin-down electrons on the bottom.

\begin{figure}[htb]
\centering
\begin{minipage}{\linewidth}
\centering
\includegraphics[width=\linewidth]{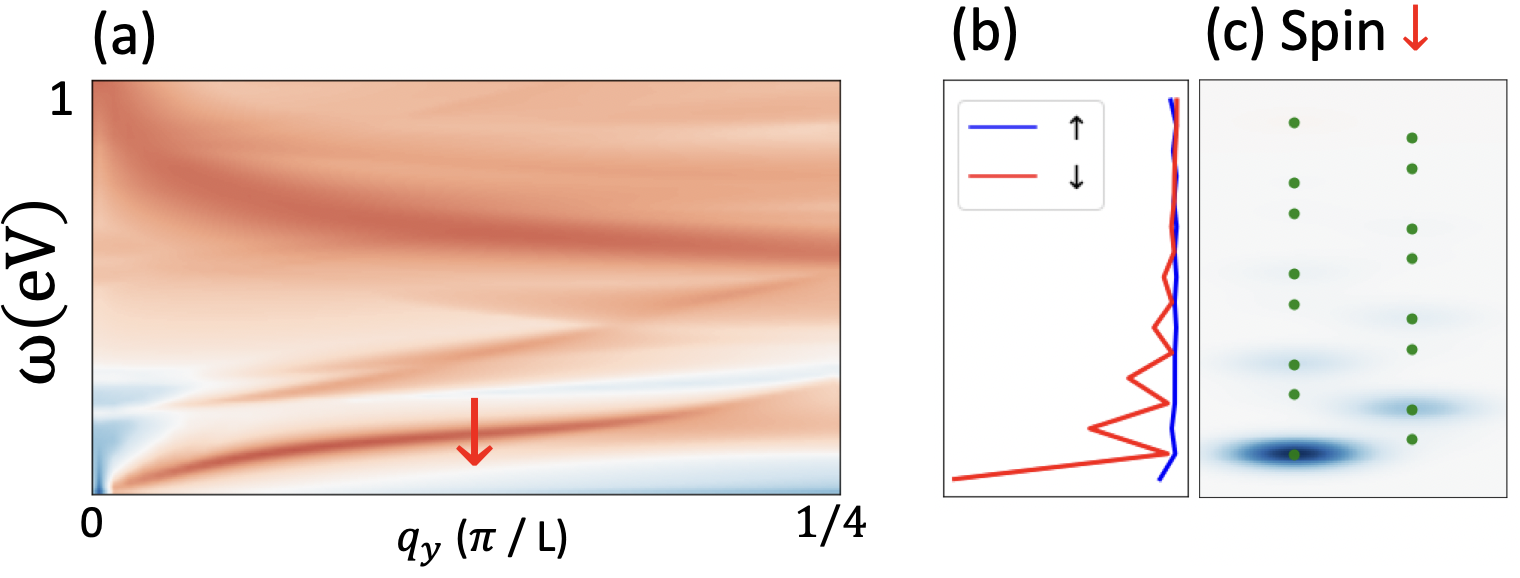}
\end{minipage}
\caption{(a) EELS of the Kane-Mele ribbon structure with a shifted Fermi level $E_F=-0.15t$ and in the presence of an in-plane magnetic field $\mathbf{B_x}=0.15\,t\,\mathbf{\hat{x}}$. Gapless edge plasmons are observed for spin-down component only. (b) Spin projection and (c) real-space charge modulation pattern of the plasmon mode marked by the red down-arrow in (a).}
\label{fig:Bx_plasmon}
\end{figure}

To illustrate this, we calculate the EELS of the Kane-Mele ribbon in presence of a $\mathbf{B_x}$ field with the Fermi level set to be $E_F=-0.15t$ (crossing the dashed red band). The result is plotted in Fig.~\ref{fig:Bx_plasmon}(a), from which we can clearly see a gapless plasmon dispersion branch. As expected before, these plasmons are of spin-down electrons localized on the bottom edge of the ribbon. For instance, we show the real-space charge modulation pattern of the plasmon mode at $q_y=\pi/8L$ (indicated by the red down-arrow) in Fig.~\ref{fig:Bx_plasmon}(c). This plasmon is indeed an edge mode localized on the bottom edge. Moreover, we project this charge distribution onto each spin component and show them in Fig.~\ref{fig:Bx_plasmon}(b). We observe that the mode is strongly spin-polarized, whereby the spin-down component dominates. It is worth mentioning that in previous sections of non-gapped models (no matter with $\mathbf{B_z}$ field or not), gapless edge plasmons always come in pairs of spin-up component and spin-down component (see Fig.~\ref{fig:KM_EELS}(b) and Fig.~\ref{fig:Zeeman_band}). Although spin-up plasmons and spin-down plasmons are excited on different edges, they are energetically degenerate. In the present case, the phenomenon is fundamentally different, as we only have one spin component of gapless edge plasmons excited. However, it is fair to clarify that the realization of selective excitation of spin-plasmons here is at the cost of losing chirality. From the electronic structure in Fig.~\ref{fig:Bx_band}, we can see that the back-scattering channel within the conducting dashed red band is open. Therefore, spin-down plasmons localized on the bottom edge with negative $q_y$ can also be excited. This is distinct from the situation with $\mathbf{B_z}$, where we have observed non-reciprocity.

\subsection{Real-space edge plasmons in Kane-Mele model on a diamond-shaped nanoflake}
\begin{figure}[htb]
\centering
\begin{minipage}{\linewidth}
\centering
\includegraphics[width=\linewidth]{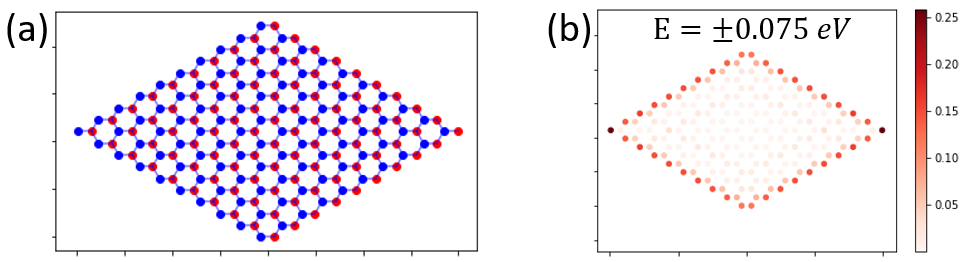}
\end{minipage}
\caption{(a) Structure of the Kane-Mele model on a diamond-shaped honeycomb lattice nanoflake with zigzag edges. (b) Modular square of degenerate single-electron edge states of the model at energies $\pm 0.075\text{ eV}$. }
\label{fig:diamond structure}
\end{figure}

\begin{figure}[htb]
\centering
\begin{minipage}{\linewidth}
\centering
\includegraphics[width=\linewidth]{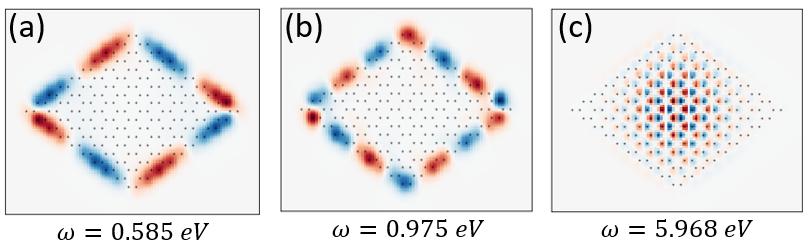}
\end{minipage}
\caption{Real-space charge modulation patterns of typical plasmon modes in the QSH Kane-Mele model on a diamond-shaped nanoflake. (a) and (b) are low-energy edge plasmons excited at $\omega=0.585$ eV and $\omega=0.975$ eV. (c) is a high-energy bulk plasmon excited at $\omega=5.968$ eV.}
\label{fig:diamond_plasma}
\end{figure}

In this section, we study plasmons in the Kane-Mele model on a different structure - a diamond-shaped nanoflake of honeycomb lattice with open boundaries of zigzag termination. This structure is illustrated in Fig.\ref{fig:diamond structure}(a) with nearest neighbour bond length of $a=1.42\AA$. We study the model in the QSH phase parameterized with $t_{\mathrm{SO}}=0.2\,t$ and $t_v = 0$. In this case, we find low-energy edge states shown in Fig.\ref{fig:diamond structure}(b). Next, we study the plasmonic excitations in the model using the full real-space RPA method introduced in \cite{gjh21}. The system supports both low-energy edge plasmons and high-energy bulk plasmons. We show three typical modes in Fig.~\ref{fig:diamond_plasma}. In the Figs.~\ref{fig:diamond_plasma}(a) and (b) we see two edge modes with different energies and wave-vectors. This indicate that low-energy edge plasmons are dispersive. The edge plasmons are extremely localized along the boundary of the nanoflake. A typical bulk plasmon mode at high energy is shown in Fig.~\ref{fig:diamond_plasma}(c) with charges oscillating inside the material.

\section{Conclusions}\label{sec:conclusion}

In conclusion, we have performed a detailed study on plasmons in the Kane-Mele model with theoretical analysis and numerical calculations. Plasmonic excitations are identified from the electron energy loss spectrum derived from the dielectric function calculated within the random phase approximation (RPA). We have mainly considered Kane-Mele model on a ribbon structure in the topologically non-trivial QSH phase where spin-polarized gapless edge plasmons have been observed. For each spin-polarization plasmons propagate uni-directionally on the edges of the ribbon. On the same edge, spin-up plasmons and spin-down plasmons always propagate in opposite directions. For a specific momentum transfer $q_y$ along the longitudinal direction of the ribbon, both spin-polarized plasmons can be excited, but localized on opposite edges. We summarize these three properties as \textit{chirality}, \textit{spin-momentum-locking} and \textit{spin-edge-locking} of plasmons in QSH insulators. Plasmons in QSH phase are sensitive to external magnetic field. Applying an out-of-plane magnetic field $\mathbf{B_z}$ breaks the TR symmetry by effectively magnetizing the material. However, it does not open a gap without introducing any non-TR-invariant spin-mixing term. Therefore, spin-polarized gapless edge plasmons remain, but the dispersion spectrum is no longer TR symmetric. We have found that increasing the field strength of $\mathbf{B_z}$ will gradually delocalize edge plasmons in one direction, but does not significantly affect edge plasmons in the opposite direction. On the other hand, applying an in-plane magnetic field $\mathbf{B_x}$ along the transverse direction of the ribbon opens a gap on each edge by breaking the TR symmetry with onsite spin-flip term. In this case, spin-up and spin-down electron bands on the same edge are energetically shifted. By tuning the chemical potential via proper doping or gating, we can selectively make one spin-polarized edge band conducting and therefore excite gapless edge plasmons with specific spin-polarization. However, we have clarified that these edge plasmons are no longer chiral due to opened back-scattering channel. Our results imply that spin-polarized plasmons in 2D QSH topological insulators can be manipulated by external magnetic field. Besides the ribbon structure, we have also investigated plasmons of the QSH Kane-Mele model on a diamond-shaped nanoflake with open boundaries. We have observed low-energy edge plasmons circulating the sample, which can also be interpreted as a topological signature of the model in its in collective excitations.  

We would expect spin-polarized plasmons studied in our work could possibly be detected by magneto-optic Kerr effect (MOKE) in experiments~\cite{Wang2021}. Moreover, non-reciprocity of edge spin-plasmons induced by $\mathbf{B_z}$ field may find potential applications in novel plasmonic or spintronic devices. In our future work, we want to study robustness of the gapless edge spin-plasmons against perturbations, such as a non-zero Rashba spin-mixing coupling or localized impurities on the edge~\cite{gjh21}. Due to the spin-edge-locking of edge plasmons, We are interested in engineering spin-plasmons on different local edges via inhomogeneous substrate screening ~\cite{Jiang2021}. Moreover, we are also interested in further exploring  plasmons, or other types of electronic collective excitations, in a three-dimensional topological system, such as Bernevig-Hughes-Zhang (BHZ) model~\cite{bhz06}.

\begin{acknowledgments}
We wish to acknowledge useful discussions with Chunyu Tan. Zhihao Jiang acknowledges the US Department of Energy, Office of Science, Materials Science and Engineering Division (under Contract No. DE-SC0022060) for supporting part of the manuscript preparation.
\end{acknowledgments}

\appendix

\section{Topological origin of gapless edge plasmons in Kane-Mele ribbon structure}\label{app:Topo_origin}

\begin{figure}[htb]
\centering
\begin{minipage}{\linewidth}
\centering
\includegraphics[width=\linewidth]{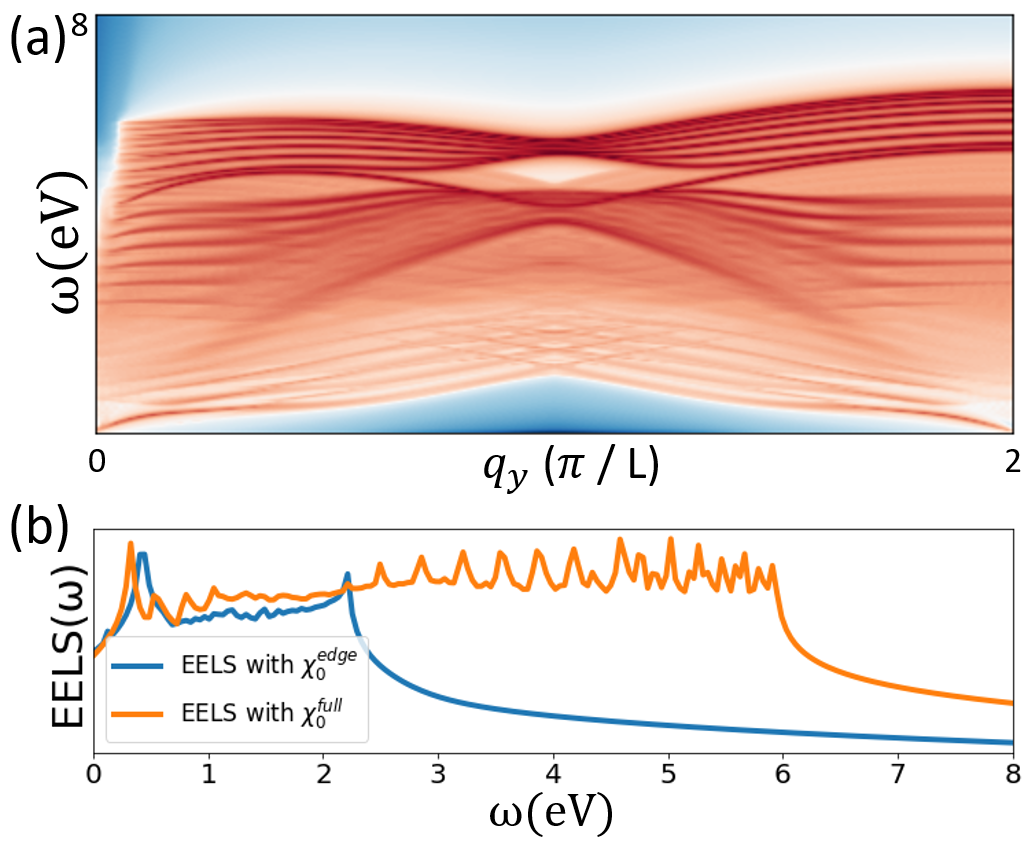}
\end{minipage}
\caption{(a) Plasmon dispersions of the Kane-Mele ribbon in the QSH phase. (b) Electron energy loss spectrum, i.e. $\text{EELS}(q_y=\pi/8L, \omega)$, of the Kane-Mele ribbon in the QSH phase evaluated with full susceptibility $\chi_0^\text{full}$ and edge bands susceptibility $\chi_0^\text{edge}$, respectively.}
\label{fig:KM_EELS_edge_bulk}
\end{figure}

We have observed high-energy bulk plasmons as well as gapless edge plasmons in the QSH phase of the Kane-Mele ribbon structure from the dispersions shown in Fig.~\ref{fig:KM_EELS_edge_bulk}(a). Here we demonstrate that these gapless edge plasmons have topological origin in the sense that they are directly originated from topological conducting edge states in the single-electron band structure. To this end, we apply the method of decomposing the full polarization function, denoted as $\chi_0^{\text{full}}$, into the edge part (involving transitions only within conducting edge bands) denoted as $\chi_0^\text{edge}$ and the rest part involving other bulk bands. Such a decomposition has been introduced in~\cite{gjh21} in order to disentangle the bulk-states and edge-states contributions to plasmonic excitations in a topologically non-trivial system. In Fig.~\ref{fig:KM_EELS_edge_bulk}(b) we show the $\text{EELS}(\omega, q_y)$ for the QSH phase at a small momentum transfer $q_y = \pi/\,8\text{L}$, using $\chi_0^{\text{full}}$ and $\chi_0^{\text{edge}}$ respectively. Comparing both spectra, we can see that $\chi_0^\text{edge}$ qualitatively reproduces the low-energy part ($\omega\,<\,2.5\,\text{eV}$) of the full spectrum with the full polarization $\chi_0^\text{full}$ considered. This indicates that the low-energy plasmonic excitations are dominated by electronic edge states. As we have already confirmed in the main text, both conducting electronic states and gapless plasmon modes are strongly localized along the edge of the ribbon. Therefore, gapless edge plasmons indeed have topological origin.

\section{Plasmons in a large-bulk-gap QSH Kane-Mele ribbon with external magnetic field $\mathbf{B_z}$} \label{app:large_gap_Bz}

\begin{figure}[htb]
\centering
\begin{minipage}{\linewidth}
\centering
\includegraphics[width=\linewidth]{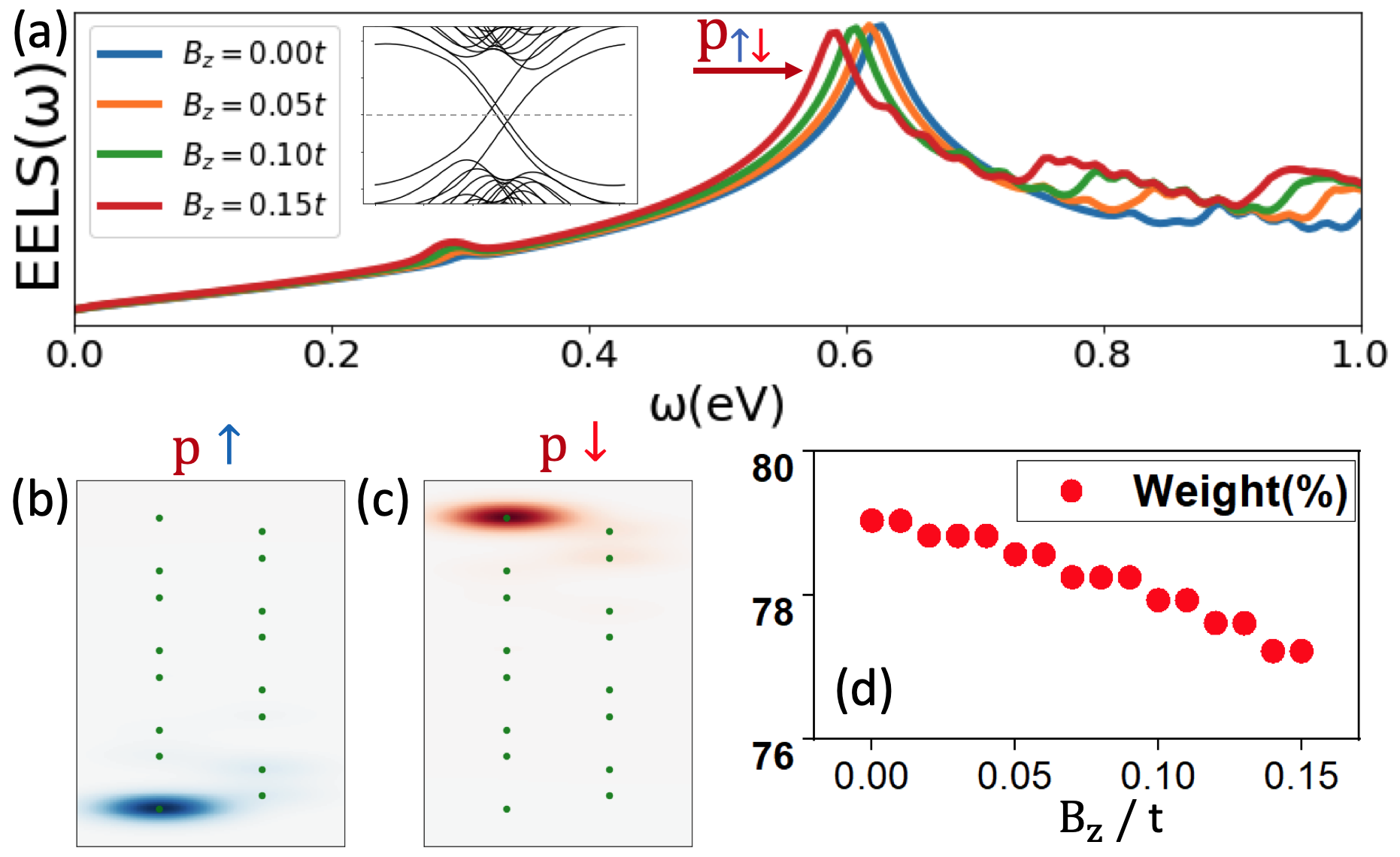}
\end{minipage}
\caption{(a) EELS of the QSH Kane-Mele model along the momentum transfer $q_y = \pi/8L$ in the presence of external magnetic field $\mathbf{B_z}$ of different strengths. Spin-orbit interaction in the numerical calculation here is $t_{\mathrm{SO}}=0.15\,t$. The inset shows the electronic band structure with $B_z=0.15t$. (b) and (c) are real-space charge modulation patterns of two spin-plasmons excited at ``P" with $B_z=0.15t$. They are still strongly localized on edges. (d) shows a slightly decreasing weight of charges on the edge for the plasmon mode ``P" with increased strength of the field.}
\label{fig:Bz_EELS_singkleK_big_bulk}
\end{figure}

We parameterize a QSH Kane-Mele ribbon with a large bulk band gap by increasing the diagonal spin-orbit coupling to $t_{\mathrm{SO}}=0.15t$ (Eq.~\eqref{Eq:Hamiltonian}). This makes bulk energy bands to be further away from conducting edge states at the Fermi level. In this scenario, a small external magnetic field $\mathbf{B_z}$ will shift energy bands, but will not significantly affect the edge plasmon excitation and real-space localization. In Fig.~\ref{fig:Bz_EELS_singkleK_big_bulk}(a) we show the EELS along $q_y=\pi/8L$ (same as the one in the main text) for different strength of the magnetic field $\mathbf{B_z}$. As we can see, the edge plasmon excitation energy does not change very much, and its real-space charge modulation patterns for both spin components are still strongly localized at the edges of the ribbon. A quantitative description of the edge charge weight with increasing magnetic field up to $B_z=0.15t$ is shown in Fig.~\ref{fig:Bz_EELS_singkleK_big_bulk}(d). We can see the maximum change is only about $2\%$. In this situation, edge plasmons are quite insensitve to the external magnetic field $\mathbf{B_z}$.        


\section{Plasmonic excitations in the graphene-like honeycomb lattice}\label{sec:Graphene}
\begin{figure}[htb]
\centering
\begin{minipage}{\linewidth}
\centering
\includegraphics[width=\linewidth]{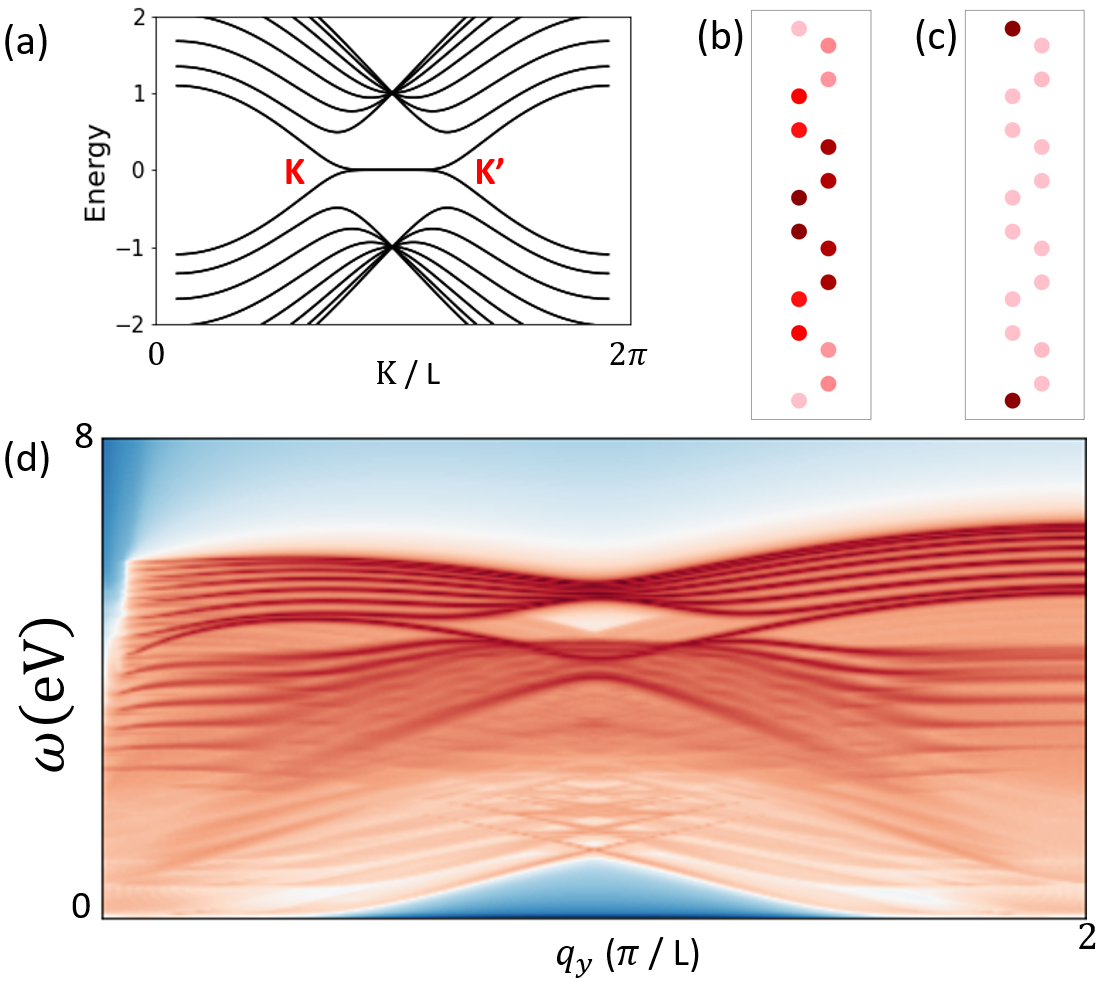}
\end{minipage}
\caption{(a): Single particle energy spectrum of the tight-binding nano-ribbon with zigzag edges. (b)-(c): Probability density of two wave functions (\text{$|\psi|^2$}) at momentum $K=\pi/L$. (d): Electron energy loss spectrum (EELS) of the ribbon structure in the first Brillouin zone with periodic boundary in the X direction (parallel to the edges) and open boundaries in the Y direction (perpendicular to edges).}
\label{fig:Graphene}
\end{figure}

Instead of only discussing topological insulators, we analyze plasmons in a graphene-like model with a simple zigzag nano-ribbon (ZNR) structure as a compared benchmark, considering the tight-binding Hamiltonian,
\begin{equation}
\hat{H} = t\sum_{<i,j>}c_i^+c_j,
\end{equation}
where $<i,j>$ represents nearest neighbor sites on the honeycomb lattice, and $t$ is the corresponding hopping parameter. In Fig.\ref{fig:Graphene} we show results for a nanoribbon with 16 atoms in the unit cell. We first inspect the single particle energy bands, shown in Fig.\ref{fig:Graphene}(a), and  wave functions at particular momenta, shown in Figs.\ref{fig:Graphene}(b) and (c). The finite size of the nanoribbon produces confinement of the electronic states in the regions near the Dirac points. The two bands in the gap overlap to form a state around $K/L=\pi$. By plotting the combined wave functions, we  find that these zero-energy states in Fig.\ref{fig:Graphene}(c) localize at the edges of the strip-like unit cell, whereas the other, higher energy, wave functions at the same momentum are confined to the bulk. The two bands of the localized edge states that occur between $\textbf{K}$ and $\textbf{K'}$ in Fig.\ref{fig:Graphene}(a) are  affected twofold by the finite width, i.e., they merge and are slightly offset from zero\cite{bf06}.    

We now examine plasmonic excitations in this basic model and calculate the EELS along the Y-direction in momentum space while accounting for the spin degeneracy. Fig.\ref{fig:Graphene}(d) shows the EELS in the first Brillouin zone, where we observe  clusters of plasmonic branches in different energy regions. At high energies ($\omega> 5 \,\text{eV}$), there are 8 plasmonic bands  corresponding to the  8 A-sites in the unit cell. Similarly, there are also 8 plasmonic bands at intermediate energies ($\omega \in\{2\,\text{eV}, 5\,\text{eV}\}$), including an arrow-shaped band at the exact center of the first Brillouin zone. In the low energy region ($\omega< 2 \,\text{eV}$), we  observe a continuum of excitations as well as several quasi-bands. By analyzing the single particle energy structure  and the EELS of the ribbon, we  conclude that the bulk energy bands and the geometry of the model are the original source of the high and intermediate energy plasmonic excitations. In contrast, the low energy continuum is due to the edge modes. 

\nocite{*}
\emergencystretch=2em
\bibliography{ref, ref_zj}
\end{document}